\documentclass[prl,superscriptaddress,preprint]{revtex4}

\usepackage{graphicx}
\usepackage{amssymb}
\usepackage{amsmath}
\usepackage{amsfonts}
\usepackage{epstopdf}
\usepackage{epsfig}
\usepackage{wrapfig}
\usepackage{color}
\usepackage{mathrsfs}

\begin{document}

\title{Work and energy gain of heat-pumped quantized amplifiers}

\author{D. Gelbwaser-Klimovsky}
\affiliation{Weizmann Institute of Science, 76100
Rehovot, Israel}
\author{R. Alicki}
\affiliation{Weizmann Institute of Science, 76100
Rehovot, Israel}
\affiliation{Institute of Theoretical physics and Astrophysics,
University of Gda\'nsk}
\author{G. Kurizki}
\affiliation{Weizmann Institute of Science, 76100
Rehovot, Israel}

\begin{abstract}
We investigate heat-pumped single-mode amplifiers of quantized fields  in high-Q cavities  based on \textit{non-inverted} two-level systems. Their  power generation is shown to crucially depend on the capacity of the quantum state of the field  to accumulate useful work. By contrast, the   energy gain of the field is shown to be insensitive to its quantum state. Analogies and differences with masers are explored.

\end{abstract}
\maketitle

\textbf{Introduction}
Thermodynamics divides the energy interchanged between systems and their surrounding (``baths'') into heat and work in an asymmetric fashion: While work can be entirely transformed into heat in a cyclic process, the opposite is not true\cite{CallenBOOK85}: the   energy  exchanged  by a system  in a closed cycle is divided between  work, $W$, and heat, $Q$. For a quantum system  the    accepted  division is given by \cite{GemmerBOOK10,AlickiJPA79,GeusicPR67}

\begin{equation}
W=-\int_{cycle} tr(\rho_S\dot{H}_S) dt;
\quad Q=\int_{cycle} tr(\dot{\rho_S}H_S) dt.
\label{eq:stwork}
\end{equation}

\noindent Both $W$ and $Q$ are determined by the parametrically (\textit{classically}) driven  Hamiltonian $H_S(t)$ and by $\rho_S(t)$  the reduced density matrix of the system S. Positive $W$ stands for work done in a cycle by S.
Equation \eqref{eq:stwork} underlies the numerous  quantum heat engine (QHE) models that have been proposed to date \cite{GemmerBOOK10,AlickiJPA79, kosloffarxive13,GeusicPR67,FordPRL06,GevaJMO02,QuanPRE07,VandenBroeckPRL05,LinPRE03,EspositoPRL09, KieuPRL04, LindenPRL10,agarwalarxiv13, ScullyNAS11,BlickleNATP11,AllahverdyanPRE05,campisiPRL09,campRMP11}. These models  may be deemed \textit{semiclassical}, as they employ classical fields or forces that drive quantum systems \cite{ScullyBOOK97}.  In these models the cycle usually  consists of ``strokes'' (legs) in which the system (``working fluid''), S, alternates between coupling  to the ``hot'' (H) and ``cold'' (C) heat baths.  

We have recently \cite{Gelbwaserumach,KolarPRL12} put forward  a \textit{minimal} semiclassical QHE model whose ``working fluid''   consists of a  two-level system  (TLS) that is \textit{permanently} coupled to \textit{spectrally separated} H and C baths and is governed by

\begin{equation}
H_S(t)=\omega_0(t)\sigma_Z, \quad \omega_0(t)=\omega_0(t+\tau). 
\label{eq:Hs}
\end{equation}

\noindent Here $\sigma_Z$ is the population-difference operator and $\tau$ is the period of the parametric modulation (Stark-shift) effected, e.g., by an off-resonant classical field with sinusoidal amplitude $E_0sin(\frac{\pi t}{\tau}):$ the modulating field plays the  role of a ``piston'' (P).
This nonadiabatic model  circumvents the  problem of \textit{breaking a finite-time cycle into strokes}: the commonly assumed  abrupt on-off switching of system-bath interactions in separate strokes of such nonadiabatic cycles  may strongly affect their  energy and entropy exchange and thereby their quantum state, which casts doubts on the validity of existing models of finite-time engines, as they ignore such effects. 

Yet what if we wish to analyze the performance of \textit{fully quantized }(rather than semiclassical) engines \cite{Skrzypczykarxive13}? Namely, what if  we wish to examine the power generation in a QHE  upon preparing the \textit{initial quantum state} of the piston (field) P, then coupling P to the system S and letting S+P evolve under the \textit{time-independent} Hamiltonian $H_{S+P}$ along with the permanent action of the baths H and C? Such scenarios may arise  in optomechanical  and nanomechanical setups\cite{xiangRMP13,Aspelmeyerarxiv13} or in cavity-field amplifiers (masers)\cite{ScullyBOOK97,ScovilPRL59}. Understanding  such scenarios is not only important for applied purposes but  is  a necessary step towards   clarifying the  fundamental question: is  thermodynamic performance different for quantum and classical or semiclassical devices?
Here we   explore  this question for QHEs, by asking specifically: 
a) What is the proper definition of work   when the standard formula (Eq. \eqref{eq:stwork}) \textit{does not apply},  since  P   affects  S via a \textit{time-independent} Hamiltonian? To this end, we invoke a seldom-used but  rigorous definition of work in fully quantized setups  \cite{LenardJSP78,puszcmp1978}.
b) Is work extraction  sensitive to the   quantum state of P, and is work distinguishable from energy exchange in a QHE? We investigate this virtually \textit{unexplored} issue  in  a  quantized analog of the semiclassical  minimal model described by Eq. \eqref{eq:Hs}, i.e., to a  heat-pumped cavity-field amplifier   and compare it to its known maser  predecessors \cite{ScullyBOOK97,ScovilPRL59}. Work fluctuations \cite{JarzynskiCRP07} will be considered elsewhere.

\textbf{Work in fully quantized setups}
Work extraction from a QHE  is governed by a non-unitary Liouville  equation, which accounts  for the  \textit{indirect} interaction of P with the baths,  mediated by  S. One may be tempted to exclude  S and consider  P as the working fluid, but such a QHE would not  be  autonomous, i.e., it would require external parametric driving. What determines the maximum work extractable by P from the baths via S?
If P  could \textit{unitarily} evolve from its state $\rho_P$ to another state, then the \textit{energy difference} between the two states would be   purely work, without entropy or heat change.
Hence,  for a given $\rho_P$ the maximum  extractable work  is

\begin{equation}
W_{Max} (\rho_P)= \langle H_P(\rho_P) \rangle -\langle H_P(\tilde{\rho}_P) \rangle
\label{eq:maxw}
\end{equation}

\noindent where  $\tilde{\rho}_P$ is a unitarily-accessible  state (i.e. a state with the same entropy as $\rho_P$) that minimizes the mean energy of P, $\langle H_P \rangle$ (and thus \textit{maximizes}  the  work extraction), known as a passive state \cite{LenardJSP78,puszcmp1978}.  By this definition, for passive states $W_{Max}(\tilde{\rho}_P)=0$ (no extractable work).

The  upper bound on $W_{Max} (\rho_P)$  is obtained  by minimizing the second term on the r.h.s. of  Eq. \eqref{eq:maxw}.
This would ideally  require  a unitary transformation of $\rho_P \rightarrow \tilde{\rho}_P $ such that $
\langle H_P (\tilde{\rho}_P) \rangle = \langle H_P  \rangle_{Gibbs}$, since  the passive Gibbs state 
 is the minimal-energy state with the same entropy as $\rho_P$ and a \textit{real} temperature $T_P$:  $(\tilde{\rho}_P)_{Gibbs}=Z^{-1} e^{-\frac{H_P}{T_P}},
$
where $Z$ is a normalizing factor.
Yet such a unitary process typically does \textit{not exist}. We therefore extend the definition \cite{LenardJSP78,puszcmp1978} in \eqref{eq:maxw} as follows: we  parametrize  the unitarily (or iso-entropically)  accessible $\tilde{\rho}_P$ with the least $\langle H_P \rangle$ as if it were a Gibbs state, i.e. $T_P$ is defined as an \textit{effective} temperature.

 Eq. \eqref{eq:maxw} may be adapted to  \textit{nonunitary} engine action: as the initial piston state $\rho_P^{(i)}$ evolves    to  a new state $\rho_P^{(f)}$, the corresponding  \textit{change} of the maximum extractable work can be expressed as  $\Delta W_{Max}=W_{Max} (\rho^{(f)}_P)-W_{Max}(\rho^{(i)}_P)$, since Eq. \eqref{eq:maxw} \textit{defines} $W_{Max}$ for each state, and may be used to infer $\Delta W_{Max}$ as $\rho_P^{(i)}\rightarrow\rho_P^{(f)}.$
Upon taking the time derivative of this bound, we find that the  extractable power in a nonpassive state of P  has the upper bound

\begin{gather}
\mathcal{P}^{Max}_{nonpas}=
\frac{d}{dt} \langle H_P (\rho_P) \rangle -T_P \dot{\mathcal{S}}_P,
\label{eq:pnonpas}
\end{gather}

\noindent where we have used the fact that the entropy rate of change of the effective Gibbs state satisfies $T_P \frac{d\mathcal{S}_P}{dt}=\frac{d\langle H_P\rangle_{Gibbs}}{dt} $.

To compare this bound on power to its semiclassical counterpart, we may  invoke the first law of thermodynamics (energy conservation) for the total (closed) complex of S, P, cold (C) and hot (H) baths. In doing so we  assume that S has reached the \textit{steady-state}, so that $ \langle H_{S}\rangle$ and the entropy of S have become constant. The standard  expression for power generated by the QHE (also known as the canonical expression) has then the form \cite{AlickiJPA79}

\begin{equation}
\frac{d \langle H_{P}(\rho_P)\rangle}{dt}=\mathscr{J}_C+\mathscr{J}_H,
\label{eq:can}
\end{equation}

\noindent where the heat currents $\mathscr{J}_C$ and $\mathscr{J}_H$ express the energy flow between the  respective baths and S \cite{GevaJMO02,AlickiJPA79,Gelbwaserumach,KolarPRL12}.  Equation \eqref{eq:can} follows from  the first law of thermodynamics, but the customary identification of the engine power with $\frac{d \langle H_{P}(\rho_P)\rangle}{dt}$ may be untrue: such identification  \textit{ignores the entropy change of P} with time, which may only be justified if P is a \textit{classical} field. Consequently, the use of $\frac{d \langle H_{P}(\rho_P)\rangle}{dt}$  fails to distinguish between work and heat production in the quantum limit, as opposed to Eq. \eqref{eq:pnonpas}.

\begin{figure}
	\centering
		\includegraphics[width=0.5\textwidth]{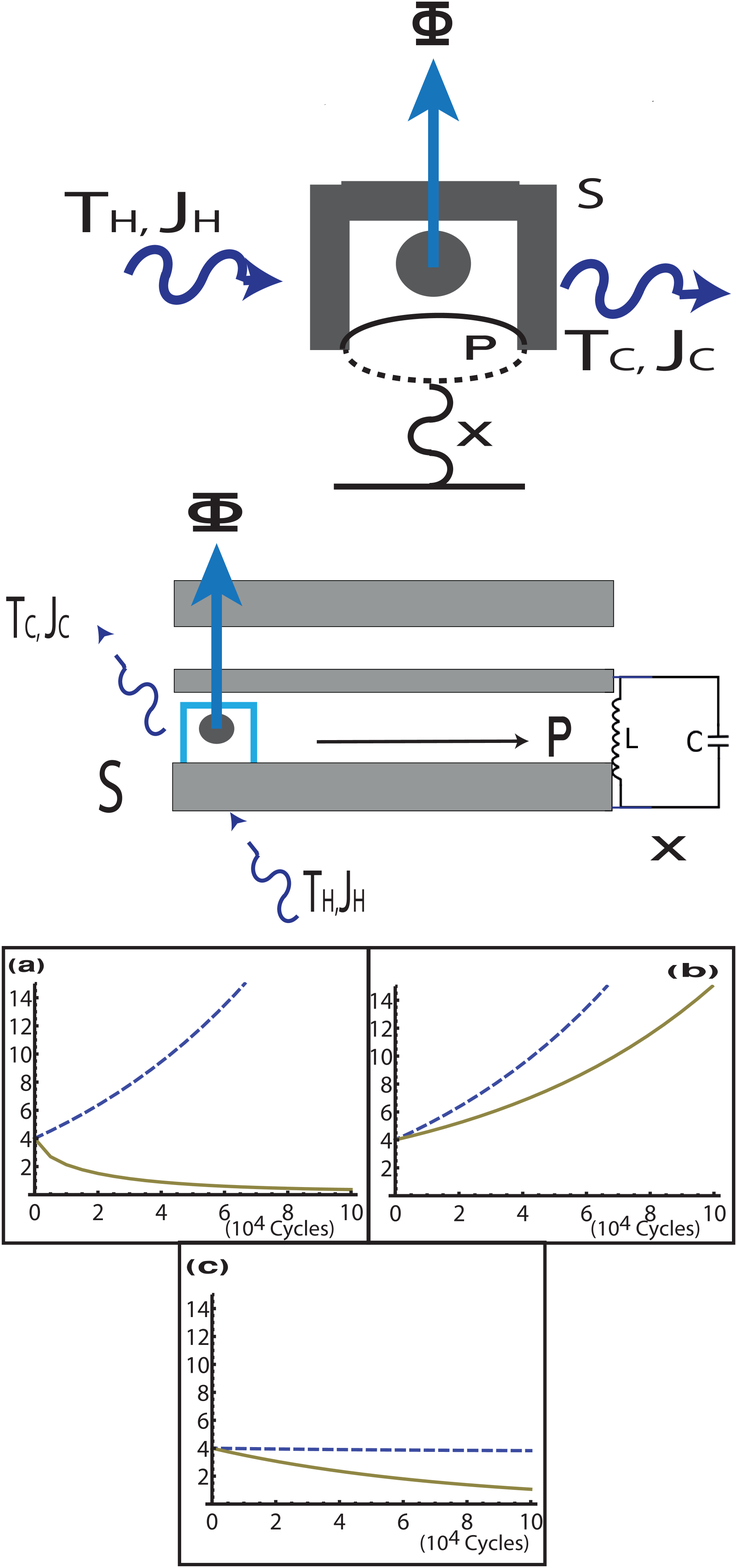}
	\caption{Upper  inset : A superconducting flux qubit S coupled to baths H,C. The piston P is a  mechanical oscillator (cantilever) whose position affects $\Phi$ and thereby the energy of S. P has the capacity to do  mechanical work on external oscillator X. Lower inset: Analogous setup, where P is a coplanar-resonator mode, whose magnetic field contributes to the flux $\Phi$ through the qubit. Here P has the capacity to do electromagnetic work on LC-circuit X. Main panel: Energy gain (dashed) and work-capacity bound (solid)    for an initial (a) Fock state, (b) coherent state, in the gain regime $\gamma=-1.39*10^{-4}$ (energy-arbitrary units, cycles=$\nu t=2\pi$). (c) For $\gamma=1.39*10^{-4}$ (loss regime) or single bath,  the coherent-state work capacity decreases rapidly while the energy decreases slowly.  The value of $\gamma$ does not affect the bound but work extraction demands $\gamma <0$. }
	\label{fig:pdistribution2}
\end{figure}

 What is required for an \textit{increase} in the maximum extractable work $W_{Max}$  as the QHE evolves, i.e,  for $\mathcal{P}^{Max}_{nonpas} >0$? We find that the necessary (but not sufficient) condition  is that the initial state  $\rho_P^{(i)}$ \textit{must  be  nonpassive,}   since   (dissipative or amplifying)  evolution   may not change a passive state to a nonpassive one, at least under  the standard Markovian assumption   (see below). By contrast, a \textit{nonpassive}  P  state  under such evolution may yield $\Delta W_{Max}>0$, i.e., it allows   potential work  \textit{accumulation} in P.  This accumulation (capacity) can be turned into real (extracted) work if and when P is coupled  to an \textit{external} degree of freedom (X);  e.g., a  piston in an opto-(or nano)-mechanical setup \cite{xiangRMP13,Aspelmeyerarxiv13} can be coupled to an external cantilever (Fig.\ref{fig:pdistribution2}-inset). The amount of extracted work will depend on the X-P coupling, but the capacity can be evaluated \textit{independently} of this coupling.

\textbf{Cavity-based heat-pumped amplifier}  The implications  of the general work and power analysis  presented  above will be elucidated for  the \textit{simplest} (\textit{minimal}) model  of an \textit{autonomous} QHE: a harmonic-oscillator P (the simplest \textit{unbounded-energy} piston) that acts \textit{dispersively} on a \textit{qubit} S (i.e. without changing its level populations), while S  is permanently coupled to  baths H, C. This quantized analog of the  semiclassical  model of Eq. \eqref{eq:Hs}   \cite{Gelbwaserumach} will be shown to allow heat-pumped amplification of P via \textit{non-inverted} S.
 At the heart of this  model is the off-resonant coupling of P to the $\sigma_Z$ operator of S:  

\begin{equation}
H_{SP}=g(a+a^{\dagger})\sigma_Z
\label{eq:Hsp}
\end{equation}

\noindent where $g$ is the coupling strength and $a$ and $a^{\dagger}$ are respectively  the P-mode annihilation and creation operators.  This coupling is realizable,  e.g., in the dispersive regime of  a superconducting qubit in a  resonator (see below) (Fig. \ref{fig:pdistribution2}-inset) \cite{blaisPRA04,xiangRMP13}.

The engine  is fueled by the H bath.  The system-bath  (S-B) coupling  has the  spin-boson form $H_{SB}=\sigma_X(B_H+B_C)$ where $B_{H(C)}$ are the multimode bath operators. Namely, the dipolar  $|1\rangle-|0\rangle$ transition  between the qubit states couples to two mode-continua  at different temperatures.
Under the bath-induced dynamics for the S+P state, $\rho_{S+P}(t)$, S  \textit{reaches  the steady state while  P still slowly gains (or loses) energy and work} over many cycles. From the reduced density matrix of the piston $\rho_P=Tr_S \rho_{S+P}(t)$ we can then compute the heat currents, the entropy $\mathcal{S}_P(\rho_P)$ and the effective temperature $T_P$ in Eq. \eqref{eq:pnonpas}, to find the conditions for \textit{sustainable} QHE operation.
In this analysis we wish to ensure that the second law is satisfied.
We therefore subscribe to the common description \cite{GemmerBOOK10,AlickiJPA79,kosloffarxive13,GeusicPR67,FordPRL06,GevaJMO02,QuanPRE07,VandenBroeckPRL05,LinPRE03,EspositoPRL09, KieuPRL04, LindenPRL10,agarwalarxiv13, ScullyNAS11,BlickleNATP11,AllahverdyanPRE05,campisiPRL09,campRMP11,ScullyBOOK97,Gelbwaserumach,KolarPRL12,Skrzypczykarxive13} of the bath-induced dynamics by the  Lindblad  (or, more correctly, the Lindblad-Gorini-Kossakowski-Sudarshan--LGKS) generator which adheres to the second law \cite{LindbladBOOK83}.

 In the dressed-state basis that diagonalizes the S+P Hamiltonian

\begin{equation}
a \mapsto b = U^{\dagger} a U,\ \sigma_{\pm} \mapsto \widetilde{\sigma}_{\pm} = U^{\dagger} \sigma_{\pm} U\ ,\  U=e^{\frac{g}{\nu}(a^{\dagger}-a)\sigma_{z}},
\label{dress}
\end{equation}
the Lindblad generator involves the bath response at the Hamiltonian eigenvalues: the resonant qubit (S) frequency $\omega_0$ and the combination frequencies $\omega_0 \pm \nu$, $\nu$ being the  piston (P) frequency. Namely

\begin{equation}
{\frac{d\rho_{S+P}(t)}{dt}}=\sum_{q=0,\pm1}(\mathcal{L}_{q}^C +\mathcal{L}_{q}^H)\rho_{S+P}(t);
\label{ME_gen}
\end{equation}
 \noindent where  $q=0,\pm 1$ labels $\omega_0 + q\nu$ of $S+P$, 
and the generators  associated with harmonic $q\nu$ in  the two baths are $\mathcal{L}_{q}^{j}$ ($j=C,H)$. These generators have the following form   for weak S-P coupling, $
\frac{g}{\nu} \sqrt{\langle b^{\dagger} b\rangle} << 1$:
\begin{subequations} \label{eq:gen}
\begin{gather}
\mathcal{L}_{0}^{j}\rho_{S+P} = \frac{1}{2}\Bigl\{G^j(\omega_0)\bigl([\widetilde{\sigma}_{-} \rho_{S+P}, \widetilde{\sigma}_{+}] + [\widetilde{\sigma}_{-}, \rho_{S+P} \widetilde{\sigma}_{+}]\bigr) +  \nonumber \\
G^j(-{\omega}_0)\bigl([\widetilde{\sigma}_{+} \rho_{S+P},\widetilde{\sigma}_{-}] + [\widetilde{\sigma}_{+}, \rho_{S+P} \widetilde{\sigma}_{-})\Bigr\},\\
\label{generator_loc}
\mathcal{L}_{q}^{j}\rho_{S+P} = \frac{g^2}{\nu^2}\Bigl\{G^j(\omega_0+ q\nu)\bigl([S_q \rho_{S+P}, S^{\dagger}_q] + [S_q, \rho_{S+P} S^{\dagger}_q]\bigr) \nonumber \\
 + G^j(-{\omega}_0- q\nu)\bigl([S^{\dagger}_q \rho_{S+P}, S_q] + [S^{\dagger}_q, \rho_{S+P} S_q ]\bigr)\Bigr\}\ , 
\ q=\pm1 .
\label{generator_loc1}
\end{gather}

\noindent Here the  operators in the transformed basis of \eqref{dress}  are given,  to lowest order in  $g/\nu$,  by 

\begin{gather}
\sigma_{+}(t)  \approx \nonumber 
\widetilde{\sigma}_{+} e^{i\omega_0 t}+\frac{g}{\nu}\bigl( S_{1}^{\dagger} e^{i(\omega_0 +\nu)t}- S_{-1}^{\dagger}  e^{i(\omega_0 -\nu)t}\bigr), \quad \nonumber\\
S_{1}^{\dagger}= \widetilde{\sigma}_{+}b^{\dagger}\ , S_{-1}^{\dagger} = \widetilde{\sigma}_{+} b
\label{fourier}
\end{gather}
\end{subequations}

\noindent The transition operators  $S_{\pm1}$ describe the  relaxation of S accompanied
by the  respective excitation or deexcitation of P, while $S^{\dagger}_{\pm1}$ describe their time-reversed counterparts.
The  effects  of the baths in Eqs. \eqref{ME_gen}-\eqref{generator_loc1}  are here described by  the Fourier transforms of the autocorrelation functions
$
G^j(\omega)= \int_{-\infty}^{+\infty} e^{i\omega t}\langle B^j(t)B^j\rangle_{T_j} dt = e^{\omega/ T_j} G^j(-\omega).
$

\textbf{Work extraction from a given state}
To evaluate the state-dependence of work in this model we cast the  (master) equation  for $\rho_P=Tr_S\rho_{S+P}$ (obtained from Eqs. \eqref{ME_gen}-\eqref{eq:gen}) into a Fokker-Planck (FP) equation  \cite{ScullyBOOK97,carmichaelBOOK99,Schleichbook01} for
the phase-space (quasiprobability) distribution in the coherent-state basis, $\rho_P=\int{d^2 \alpha \mathbf{P}(\alpha) |\alpha \rangle \langle \alpha|}$. 
 Upon assuming a\textit{ steady-state for S }and using standard methods \cite{ScullyBOOK97, carmichaelBOOK99}, we derive the FP equation

\begin{gather}
\frac{d \mathbf{P}}{dt}=\frac{\gamma}{2}(\frac{\partial}{\partial\alpha}\alpha+\frac{\partial}{\partial\alpha^*}\alpha^*)\mathbf{P}+D\frac{\partial^{2}\mathbf{P}}{\partial\alpha\partial\alpha^*}. \label{eq:fp}
\end{gather}

\noindent Here  the drift and diffusion coefficients, $\gamma$ and $D$, respectively, depend on $G(\omega)=\sum_{j=H,C}G^{j}(\omega)$, which is the sum of the cold- and hot-baths response spectra, sampled at   $\omega=\nu_{\pm}=\omega_0 \pm \nu$. These coefficients read

\begin{gather}
\gamma=\frac{g^{2}}{\nu^{2}}\Big((G(\nu_+)-G(\nu_-))\rho_{11}+ \notag 
(G(-\nu_-)-G(-\nu_+))\rho_{00} \Big);  \\
 D=\frac{g^{2}}{\nu^{2}}\left(\left(G(\nu_-)\rho_{11}+G(-\nu_+)\rho_{00}\right)\right)
\label{eq:gamad}
\end{gather}

\noindent Their significance becomes apparent when the  piston mean-\textit{energy (for gain or loss)} is evaluated to be 

\begin{equation}
\langle H_P(t) \rangle= \nu ( \frac{D}{\gamma} (1-e^{-\gamma t}) +e^{-\gamma t}\langle H_P(0) \rangle ).
\label{eq:energy}
\end{equation}

\noindent This mean energy is \textit{independent  of the nonpassivity }of the initial state, but is dependent on the drift of the initial energy (second r.h.s. term) countered by diffusion (first r.h.s. term). Energy gain occurs for $\gamma<0$. It   may be interpreted as heat pumping of P via   absorption of quanta by the TLS from the H bath at $\nu_+$ and their  emission to the C bath at $\nu_-$ (Fig. \ref{fig:espectro}a), endowing P with the energy $\nu_+-\nu_-=2\nu$.
Weak diffusion and high gain, $D\ll 1$ and $\gamma\ll0$, require that the two response spectra, $G^H(\omega)$ and $G^C(\omega)$ be \textit{separated} (Fig. 2a), just as in the semiclassical limit  \eqref{eq:Hs}\cite{Gelbwaserumach}.

 The  heat-pumped energy in Eq. \eqref{eq:energy}  may be partly converted into work accumulated by P, measured by  $\Delta W_{Max}=W_{Max}^{(f)}-W_{Max}^{(i)}:$  the work-capacity increase by the evolving  non-passivity of the phase-space 
distribution, as $\mathbf{P}^{(i)}(\alpha)\rightarrow \mathbf{P}^{(f)}(\alpha)$, under the FP  equation dynamics.
We ask: what determines whether a state is nonpassive and suited for
work extraction?
To  this end  we look for criteria for \textit{sustainable}  nonpassivity. This requires \textit{non-monotonic decrease} of $\mathbf{P}(\alpha)$ with the energy $\nu |\alpha|^2$ at \textit{long times}. This requirement may be formulated as follows: nonpassivity requires \textit{positive   derivative} of $\mathbf{P}(\alpha)$ with respect to  $|\alpha|$:
$
\frac{\partial \mathbf{P}(\alpha,t)}{\partial |\alpha|}>0.
$
We may use this requirement to classify work extraction by the quantum state of P, upon noting that the main contribution to the derivative comes from $|\alpha|$  close to the center of the distribution (Fig. \ref{fig:espectro}b):
 
\begin{figure}[h!]
	\centering
	\hspace{-6cm}
			\includegraphics[width=0.8\textwidth]{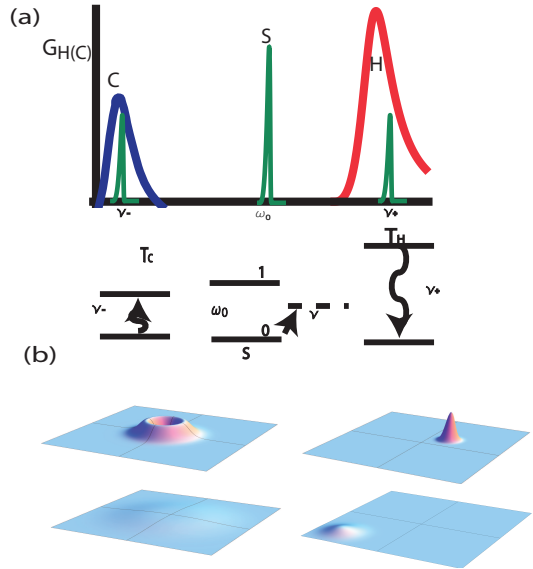}
	\caption{(a) S-B coupling spectra and level scheme. (b) Phase-plane distributions (for $\gamma<0$): left- initial  Fock state, right- Coherent state: above-- at $t=0$; below-- after $10^{4}$ cycles (same parameters as in Fig. \ref{fig:pdistribution2}. The Fock-state distribution becomes passive (thermalizes) but the coherent-state distribution increases its non-passivity.}
	\label{fig:espectro}
\end{figure}

(a) An initially passive (e.g., thermal) state will only have negative derivative, i.e.,  will remain passive at all times, with no work extraction at any time.
(b) \textit{Only $\gamma <0$}   can yield sustainable work,  since  for $\gamma>0$  (loss regime) the distribution  becomes   eventually thermal (Fig. \ref{fig:pdistribution2}c)
(c)  Since in a Fock state the center of $\mathbf{P}(\alpha)$ is at the origin ($|\alpha|=0$)  it can be shown  that the derivative eventually becomes negative and the distribution  passive  for any $\gamma$.
Thus,  a Fock state gradually loses all its nonpassivity and work capacity.
The work that may be extracted from an  \textit{initial photon-number (Fock) state} $| n^{(i)} \rangle$  is at most its initial energy $\nu n^{(i)}$: \textit{a Fock state cannot extract additional work} from  the bath-fueled engine,  hence its $\Delta W_{Max}<0$ (Fig. \ref{fig:pdistribution2}a). (d)   By contrast, in a coherent state, whose distribution is centered at $|\alpha|=|\alpha ^{(i)}|e^{-\frac{\gamma t}{2}}$  the derivative remains positive and the state can be shown to \textit{increase} its nonpassivity for $\gamma<0$, so that work extraction is sustainable at all times.
 Maximal work capacity   is then  described by a  displacement (following  the engine action) of the final distribution, $\mathbf{P}^{(f)}(\alpha)$, to $|\alpha|=0$. This displacement operation is   the unitary in Eq. \eqref{eq:maxw} that transforms $\rho_P$ to $\tilde{\rho}_P$. The coherent-state capacity  \textit{exponentially increases under  negative $\gamma$:}  $\Delta W_{Max}=\nu |\alpha^{(i)}|^2e^{-\gamma t}$ (Fig. 1b,2b)

For  a piston in an initial passive state (or even in a Fock  state), \textit{the engine  needs an ``ignition''}, which can be achieved
by \textit{coherence injection:} $|P_{ignition} \rangle =D(\alpha) |P(0) \rangle$, where $D(\alpha)$ is the displacement operator \cite{ScullyBOOK97} and $|P(0) \rangle$ is the initial piston  state.  

The fact that Eq.\eqref{eq:energy} allows the amplification of the mean energy $\langle H_P(t) \rangle$  for any initial state, even for thermal input in P is consistent with $\frac{d\langle H_P(t) \rangle}{dt}$ being the power generated according to the canonical definition \eqref{eq:can}:  Clearly, such amplified thermal field contributes to heat rather than work. Hence, the \textit{canonical definition}, as opposed to Eq. \eqref{eq:pnonpas}, does not \textit{distinguish work from heat generation.}
We have thus reached the main conclusion of this paper:  the extractable work obtainable from the nonpassivity equation \eqref{eq:pnonpas} strongly depends
on the initial state of the piston, but \textit{particularly  on its coherence, rather than purity}.

	\textbf{Experimental predictions} 
The S-P coupling in Eq. \eqref{eq:Hsp} is realizable in well-investigated experimental setups \cite{xiangRMP13,Aspelmeyerarxiv13}: 1)  A superconducting flux   qubit which is dispersively  coupled to P  may be realized by a high-Q  (phonon) mode of a nanomechanical cavity (cantilever). The quantized position of P, $a+a^{\dagger}$, affects  the magnetic flux $\Phi$  through the qubit and thereby its frequency $\omega_0$ (Fig. \ref{fig:pdistribution2}- upper inset). 2) Alternatively, P can be a field mode of a coplanar resonator \cite{blaisPRA04} whose quantized electromagnetic field position ($a+a^{\dagger}$) affects the flux  $\Phi$.  In both setups, the coupling of the qubit to a bath at $T_H$ with a spectrum centered at $\omega \sim \nu_+$ is achievable by a narrow-band \textit{local} heat-pump,  along with its coupling to a heat-dump  at $T_C$ with a spectrum limited to $\omega \leq \nu_-$ (Fig. 2a). A drift $|\gamma| \gg$ MHz is feasible. 
 In such scenarios, it should be possible to achieve $|\gamma| t_{coh} \gg 1$ where $t_{coh}\geq (\frac{\nu}{Q})^{-1}$ is the P-mode coherence time, and thus coherently  amplify the output signal. Similar considerations may apply to cold-atom \cite{PetrosyanPRA09} or spin-ensembles \cite{KuboPRL10} in high-Q cavities.

 The protocol starts  by preparing the  qubit  in the $|0>$ state,  and concurrently preparing the  piston-mode in a chosen initial state (thermal, coherent or Fock state) by established techniques \cite{ScullyBOOK97,carmichaelBOOK99}. Next, the dispersive  S-P coupling (Eq. \eqref{eq:Hsp}) is switched on by tuning their frequencies.  The main signature of the predicted effects is the dependence of work  on the initial state of P, for a given   $\gamma$,  as opposed to   the  energy gain of the P mode (Eq. \eqref{eq:energy}) that should be \textit{independent} of the initial state. Energy  gain can be measured by resonant absorption at frequency $\nu$ of phonons (Fig. \ref{fig:pdistribution2}- upper inset) or photons (Fig. \ref{fig:pdistribution2}- lower inset). Measuring  the work capacity (Fig. 1 a,b) is more subtle. We may \textit{multiplex (homodyne) the P -mode field with a local oscillator}\cite{ScullyBOOK97,carmichaelBOOK99}:  only the coherent component of the P-mode field  will then contribute to the output signal and to work, in contrast to the total energy gain which is insensitive to the coherence.

It is instructive to compare the present model with  the micromaser model \cite{ScullyBOOK97} in which the bath is at zero temperature 
and  the gain depends on the  interaction parameter  $g\tau\sqrt{\langle n\rangle }$, 
where g is the coupling between the \textit{inverted} atoms and the cavity field mode, $\langle n\rangle$ is the cavity mean photon number  and $\tau$ is the
time the atoms spend in the cavity. If the interaction parameter is assumed   small, then from Ref. \cite{ScullyBOOK97} we obtain the generated power 
$
\nu \frac{d <n>}{dt}\approx  \nu R.
$
\noindent Here $ R=r_{a}(g\tau)^{2}$, where $r_{a}$ is the
rate at which the atoms enter to the cavity .
The generated power in this case is $\nu R$. 
The invested (input) power, the energy given by the inverted atoms to the field mode, is  $\omega_{0} R=r_{a}\omega_{0}(g\tau)^{2}$. We then obtain the maximum
efficiency
$
\eta^{Max}=\frac{\nu}{\omega_{0}}
$,
i.e., the ratio between the output  and input frequency.  In our model, where atoms are \textit{noninverted}, the \textit{analogous} efficiency bound is  $\nu/\nu_+$. It is obtained from the ratio of the power output to the input heat-power $J_H$, for $|\alpha| \gg 1$, the semiclassical limit for coherent-state input.

\textbf{Conclusions}
QHE performance achievable at long but finite times\cite{GemmerBOOK10,AlickiJPA79,GeusicPR67,GevaJMO02,ScullySCI03,BlickleNATP11,AllahverdyanPRE05} upon allowing for the piston quantum state to change its entropy, is both a practically  and conceptually  important issue, here studied for piston amplification under steady-state conditions for the working fluid.
Our results  elucidate the  rapport between heat engine thermodynamics and  its hitherto unfamiliar  quantum properties. Namely, work extraction in the quantum domain  is described as \textit{deviation  from a passive state}. 
This description, extended by us to non-unitary amplification of the piston (QHE evolution), correctly divides the energy exchange into work and heat in a fully quantized setup,
by   treating the piston as a \textit{thermodynamic  resource} with effective temperature and entropy.
Remarkably,  work extraction   defined by nonpassivity crucially   depends on the initial quantum state, in contrast to energy gain that adheres to the canonical definition, which fails to distinguish work from heat production (see also \cite{goswamiPRA13}). The most striking result is that \textit{coherence rather than purity of the quantum state} determines work extraction, as manifested by the drastic difference between Fock- and coherent-state input.
Further research may strive to encompass the present scenario with that of coherence in multilevel systems \cite{ScullySCI03,BoukobzaPRA13}, entanglement in multiatom systems 
\cite{RaoPRL11,kurizkiPRA90,kurizkiPRA96,scullysci10,MazetsJPB07,Dillenschneiderepl09} and non-Markovian effects  \cite{ClausenPRL10,KofmanJMO94,ErezNAT08,JahnkeEPL10,GelbwaserPRA13}  in the general framework of nonclassical heat engines.

The support of ISF $\&$ BSF (GK),  MNiSW (RA) and CONACYT (DG) is acknowledged.

 \end{document}